\documentclass[3p,twocolumn]{elsarticle}

\usepackage{amsfonts,amsmath,amssymb}
\usepackage{graphicx,color}

\def\blfootnote{\xdef\@thefnmark{}\@footnotetext}

\begin{document}

\begin{frontmatter}

\title{Gluo-dissociation of Heavy Quarkonium in the Quark-Gluon Plasma Revisited}

\author[a]{Shile Chen }
\author[a]{Min He}

\address[a]{Department of Applied Physics, Nanjing University of Science and Technology, Nanjing 210094, China}

\date{\today}

\begin{abstract}
Using an effective Hamiltonian derived from the QCD multipole expansion, we calculate the cross sections of gluo-dissociation of heavy quarkonia in the quark-gluon plasma, by including both the chromo-electric dipole ($E_1$) as well as the chromo-magnetic dipole ($M_1$) transition mechanisms. While the former allows to reproduce the results from operator-product-expansion calculations in the Coulomb approximation, the latter as a novel contribution is shown to be significant at low energies close to the threshold. Using thus obtained cross sections, we further carry out a full calculation of the gluo-dissociation rates for various charmonia and bottomonia within a non-relativitic in-medium potential model. The $M_1$ contribution turns out to be most prominent for the $J/\psi$ and accounts for $\sim 10\%-25\%$ of the total ($E1+M1$) dissociation rate at temperatures close to the transition temperature.

\end{abstract}


\end{frontmatter}

\section{Introduction}
\label{sec_intro}
Heavy quarkonia (bound states of a heavy quark $Q$ and its anti-quark $\bar{Q}$) serve as a rich laboratory for the study of strong interactions~\cite{Brambilla:2010cs,Brambilla:2004jw}. In vacuum, a wealth of properties of these bound states can be understood in terms of a non-relativistic potential of the Cornell type consisting of a color-Coulomb term and a linear confining term~\cite{Eichten:1978tg}. When embedded into the Quark-Gluon Plasma (QGP), the properties of the heavy quarkonia thus reflect the in-medium modifications of the potential between $Q$ and $\bar{Q}$. Indeed, as the temperature of the QGP increases, the medium-induced screening of the binding force penetrates into smaller distances, resulting in the sequential dissociation of the bound states according to strength of their binding energies, as was first advocated in~\cite{Matsui:1986dk,Karsch:2005nk}.

However, on top of the static screening, inelastic collisions of the heavy quarkonia with the QGP constituents can lead to dynamical dissociation of the bound states. These processes generate inelastic widths and need to be accounted for in the description of the quarkonium in-medium spectral functions and the interpretation of experimental data~\cite{Rapp:2008tf,Rapp:2017chc,Zhao:2010nk,Zhou:2014kka,Strickland:2011mw,Song:2011nu}. The first theoretical study of inelastic collisions of heavy quarkonia with gluons could be traced back to the the seminal work by Peksin~\cite{Peskin:1979va} decades ago, as a byproduct of the operator-product-expansion (OPE) analysis of the coupling of heavy quarkonia with external soft probes. Based on the observation that the color-octet $Q\bar{Q}$ configurations can only persist over a parametrically short space-time range of the order
\begin{equation}
\Delta t\sim \frac{1}{V_8-V_1}\sim \frac{a}{g_s^2}\sim \frac{1}{\epsilon_B},
\end{equation}
(where $V_8$ and $V_1$ are the color-octet and -singlet Coulomb potential, respectively, $a$ the separation between the very massive $Q$ and $\bar{Q}$, $g_s$ the coupling and $\epsilon_B$ the binding energy of the singlet bound state), Peskin concluded that the various gluon emissions by the $Q\bar{Q}$ bound state assemble into comparably small color-singlet clusters represented by local gauge-invariant operators whose coefficients contain all the dependence on the particular heavy quarkonium. Indeed, by summing up all the diagrams, particularly including the diagrams in which the external gluons couple to the exchanged gluon between the $Q$ and $\bar{Q}$, the two-gluon emission by the $Q\bar{Q}$ bound state was shown explicitly to be described by a gauge-invariant second-order
color-electric dipole transition~\cite{Peskin:1979va}. Later on, the same problem of inelastic collision of heavy quarkonium with an external gluon was investigated by many authors~\cite{Kharzeev:1994pz,Wong:2001kn,Arleo:2001mp,Oh:2001rm, Brezinski:2011ju,Liu:2013kkg} and the leading order result (known as gluo-dissociation) by Peskin was reproduced~(the authors of~\cite{Liu:2013kkg} tried to go beyond the dipole approximation), using, e.g., the non-relativistically approximated Bethe-Salpeter amplitude for the heavy quarkonium~\cite{Oh:2001rm}. In particular, the effective field theory (EFT), built upon the hierarchies of non-relativistic and thermal scales typical of heavy quarkonium in the QGP, provides a systematic approach to address the inelastic collision of heavy quarkonium with gluons, managing to elucidate how two different inelastic mechanisms, namely, gluo-dissociation (leading order) and quasi-free (next-to-leading order) processes play their roles in different temperature regimes~\cite{Brambilla:2008cx,Brambilla:2013dpa}~(an early insight of the competition between these two mechanisms can be found in~\cite{Grandchamp:2001pf}).

The general features of Peskin's perturbative analysis could be revealed from the perspective of multipole expansion of Quantum Chromodynamics (QCD)~\cite{Gottfried:1977gp,Voloshin:1978hc,Yan:1980uh}. In particular, in Ref.~\cite{Yan:1980uh}, Peskin's perturbative analysis was promoted to the effective Langrangian level, and the complicated interactions between heavy quarks and the coupling of heavy quarks to external gluons were proved to boil down to a gauge-invariant effective action in terms of constituent quark fields. From this effective action, a non-relativistic quantum-mechanical Hamiltonian was obtained upon a gauge invariant QCD multipole expansion, appropriate to describe the coupling between the heavy quarkonium system and external soft gluons.

While the QCD multipole expansion was originally designed for the study of hadronic transitions between different heavy quarkonia~\cite{Yan:1980uh}, it provides a solid theoretical framework to address the interactions of heavy quarkonium with gluons in the QGP, where the requirement that both initial and final states be color singlets gets freed and thus transitions involving a single gluon open up. In the present work, we calculate the gluo-dissociation (the leading order inelastic collision) cross section of the charmonium and bottomonium with gluons in the QGP, using the effective Hamiltonian derived from QCD multipole expansion. In the following Sec.~\ref{sec_form}, after introducing the theoretical formalism, we first verify the correctness of our quantum-mechanical perturbation approach by reproducing Peskin's result for gluo-dissociation of $J/\psi$ due to the color-electric dipole ($E_1$) transition. Then in Sec.~\ref{sec_cross-sections} going beyond the usually adopted Coulomb approximation, we carry out a full calculation of the gluo-dissociation cross sections for various charmonia and bottomonia below threshold in a non-relativistic in-medium potential model, by including not only the color-electric dipole ($E_1$) but also the chromo-magnetic dipole ($M_1$) transitions. The latter as a novel mechanism for the gluo-dissociation of heavy quarkonium in the QGP context is shown to give a significant contribution at low energies close to the scattering threshold, as implied by the pertinent transition selection rules. In-medium cross sections thus obtained are then folded with gluon distributions in medium to yield the dissociation rates of phenomenological relevance as a function of temperature. We have found that $M_1$ contribution to the dissociation rate is most prominent for the $J/\psi$ at temperatures close to the pseudo-critical temperature for the deconfinement transition $T_{\rm pc}\simeq 170$\,MeV. Finally, we summarize and give an outlook in Sec.~\ref{sec_sum}.

\section{Gluo-dissociation from QCD Multipole Expansion}
\label{sec_form}

\subsection{QCD Multipole Expansion}
\label{ssec_Heff}
Based on the observation that the diagrams Peskin considered may be obtained by one iteration of the pertinent two-particle irreducible (2-PI) diagrams, the author of Ref.~\cite{Yan:1980uh} showed that, summation of the whole series of the all possible iterations of the 2-PI diagrams can be represented by an gauge invariant effective action in terms of the constituent quark fields and correspondingly transformed gauge fields~\cite{Yan:1980uh}, which summarizes the interactions between the heavy quarks and the couplings of the $Q\bar{Q}$ system to external gluons. Upon the multiple expansion of the changing external gauge field around the center of the $Q\bar{Q}$ system of sufficiently small size, the effective action can be transcribed into a non-relativistic quantum-mechanical Hamiltonian~\cite{Yan:1980uh},
\begin{align}\label{Heff}
H_{\rm eff}&=H_0 + H_I, \nonumber\\
H_0&=\frac{\vec p^2}{m_Q}+V_1(|\vec r|)+\sum_a\frac{\lambda^a}{2}\frac{\bar{\lambda}^a}{2}V_2(|\vec r|), \nonumber\\
H_I&=Q^aA_0^a(t,\vec 0)-\vec d^a\cdot\vec E^a(t,\vec 0)-\vec m^a\cdot\vec B^a(t,\vec 0)+...,
\end{align}
(dots in the last equation denote higher-order multipole terms) where $V_1$ and $V_2$ are the $Q\bar{Q}$ potential arising from gluon exchange ($\vec r$ being the relative $Q\bar{Q}$ separation) in color singlet and octet, respectively; together with the kinetic energy term, they make up the ``zeroth order" Hamiltonian $H_0$ of the $Q\bar{Q}$ system. The coupling of the $Q\bar{Q}$ system to the external soft gluons is represented by the interaction Hamiltonian $H_I$, in which
\begin{align}\label{multipoles}
Q^a&=g_s(\frac{\lambda^a}{2}+\frac{\bar{\lambda}^a}{2}),\nonumber\\
\vec d^a&=\frac{1}{2}g_s\vec r(\frac{\lambda^a}{2}-\frac{\bar{\lambda}^a}{2}),\nonumber\\
\vec m^a&=\frac{g_s}{2m_Q}(\frac{\lambda^a}{2}-\frac{\bar{\lambda}^a}{2})(\frac{\vec \sigma}{2}-\frac{\vec \sigma\,'}{2}),
\end{align}
are the color monopole, color-electric dipole ($E_1$) and color-magnetic dipole ($M_1$) of the $Q\bar{Q}$ system, respectively, with $\lambda^a/2$ and $\bar{\lambda}^a/2$ being the color matrices, and $\vec \sigma/2$ and $\vec \sigma\,'/2$
the spin matrices of the heavy quark and antiquark, respectively. In $H_I$, apart from the monopole term, all the other multipole interactions
are gauge invariant and are identical to their QED counterparts except for the appearance of color indices $a$. We emphasize that, to arrive at the latter
point, appropriate inclusion of the diagrams for the nonlinear self-coupling of gauge fields (i.e., coupling of the external gluons to the exchanged gluon between $Q$ and $\bar{Q}$) in the summation of perturbation series is essential.

\subsection{Deriving the Gluo-dissociation Cross Sections from $E_1$ transition}
\label{ssec_E1}

We now use the effective Hamiltonian obtained from the QCD multipole expansion to derive the gluo-dissociation cross section
of the heavy quark bound states within the quantum-mechanical perturbation approach. We use the $J/\psi$ (the vector
gound state of charm quark and antiquark bound state) as an example and illustrate the derivation of the cross section of the process: $g+J/\psi\rightarrow c+\bar{c}$  due to two different mechanisms, namely, the color-electric dipole ($E_1$) transition and color-magnetic dipole ($M_1$) transition separately.

We work with Coulomb gauge QCD
\begin{equation}
\left\{
\begin{array}{lr}
A^a_0=0, & \\
\nabla \cdot \vec A^a=0, &
\end{array}
\right.
\end{equation}
(the first condition is also known as Weyl gauge) in which the color-electric field $E^a_i=\partial_iA^a_0-\partial_0A^a_i+g_sf^{abc}A^b_iA^c_0$ reduces to
\begin{equation}
\vec E^a=-\frac{\partial \vec A^a}{\partial t}.
\end{equation}
So the interaction Hamiltonian corresponding to $E_1$ transition reads
\begin{equation}\label{H_E1}
H_{E1}=-\vec d^a\cdot \vec E^a(t,\vec 0)=\frac{g_s}{2}(\frac{\lambda^a}{2}-\frac{\bar{\lambda}^a}{2})\vec r\cdot\frac{\partial \vec A^a}{\partial t}.
\end{equation}
The gauge field can be quantized as
\begin{equation}\label{Aa}
\vec A^a(t,\vec x)=\sum_{\vec k,\lambda}N_{\vec k}\vec \epsilon_{\vec k\lambda}[a_{\vec k\lambda}^a e^{i\vec k\cdot\vec x-i\omega_{\vec k}t}+h.c.],
\end{equation}
where $\vec k$ is the gluon momentum, $\omega_{\vec k}$ the energy, $\lambda=1,2$ the two physical polarizations. With $N_{\vec k}=\sqrt{\frac{\hbar c^2}{2V\omega_{\vec k}}}$ ($V$ is the spatial volume) being the normalization constant in the rationalized Gauss unit as used here, the creation and annihilation operators of gluons satisfy the commutation relation $[a_{\vec k\lambda}^a, a_{\vec k\,'\lambda\,'}^{b\dagger}]=\delta_{\vec k\vec k\,'}\delta_{\lambda\lambda\,'}\delta^{ab}$.
Combining Eqs.~(\ref{H_E1}) and (\ref{Aa}), and using $\xi^a|\rm singlet>=(\frac{\lambda^a}{2}-\frac{\bar{\lambda}^a}{2})|\rm singlet>=\sqrt{\frac{2}{N_c}}|\rm octet,a>$ ($N_c=3$) for appropriately defined color-octet basis, one arrives at the $E_1$ transition matrix element for
the gluo-dissociation process $g(\vec k\lambda)+J/\psi\rightarrow c+\bar{c}$
\begin{equation}
H^{E_1}_{fi}=\frac{ig_s}{2}\sqrt{\frac{\omega_{\vec k}}{3V}}<(c\bar{c})_8|\vec r\cdot\vec \epsilon_{\vec k\lambda}|J/\psi>,
\end{equation}
where the gluon of momentum $\vec k$ and polarization $\lambda$ in the initial state has been annihilated and $(c\bar{c})_8$ is the final color-octet state as a result of the break-up of the initial bound state $J/\psi$.

We compute the gluo-dissociation of $J/\psi$ into a final state color-octet $(c\bar{c})_8$  in the quantum-mechanical time-dependent perturbation approach. The leading order transition rate is given by Fermi's golden rule
\begin{equation}
\Gamma^{E_1}_{i\rightarrow f}=\frac{2\pi}{\hbar}|H^{E_1}_{fi}|^2\delta(E_i-E_f).
\end{equation}
We work in the rest frame of the $J/\psi$ and neglect the interaction between the charm and anticharm quark in the final $(c\bar{c})_8$ color-octet state that has proved to be negligible for most kinematic region of interest (the largest impact of final state interaction near threshold has little effect in the dissociation rate to be calculated below; c.f.~\cite{Brambilla:2008cx}); therefore the internal motion of the final state $(c\bar{c})_8$ is represented by a plane wave of momentum $\vec p$ (relative momentum between the charm and anticharm quark). Dividing the transition rate by the flux of the incident gluon $c/V$, averaging over the polarizations and incident directions of the gluon $\frac{1}{4\pi}\int d\Omega_{\vec k}\frac{1}{2}\sum_{\lambda=1,2}$, and summing over the final state degeneracy (different momentum eigenstates of $(c\bar{c})_8$) $\sum_{\rm final~state}=\frac{V}{(2\pi)^3}\int d^3\vec p$, one arrives at an expression for the cross section
\begin{eqnarray}\label{E1crosssection}
\sigma_{E_1}^{g+J/\psi\rightarrow c+\bar{c}}(E_g)=\frac{g_s^2\pi}{2\cdot 9}E_g\frac{V}{(2\pi)^3}\int d^3\vec p \nonumber\\
\times|<(c\bar{c})_8,\vec p|\vec r|J/\psi>|^2\delta(E_g-\epsilon_B-\frac{\vec p^2}{m_Q}),
\end{eqnarray}
($\hbar=c=1$ has been used) where $E_g$ denotes the energy of the incident gluon, and $\epsilon_B$ the binding energy of the $J/\psi$.

We are now at a position to verify that Peskin's result~\cite{Peskin:1979va} can be reproduced from the above expression. For this purpose, we take the
$1S$ Couloumb bound state wave function for the $J/\psi$; the matrix element becomes
\begin{align}
&<(c\bar{c})_8,\vec p|\vec r|J/\psi>\nonumber\\
=&\int d^3\vec r\frac{1}{\sqrt{V}}e^{-i\vec p\cdot\vec r}\vec r\frac{2}{\sqrt{a^3}}e^{-r/a}Y_{00}(\theta,\phi)\nonumber\\
=&\frac{-2i}{\sqrt{Va^3}}\nabla_{\vec p}\int d^3\vec re^{-i\vec p\cdot \vec r}e^{-r/a}Y_{00}(\theta,\phi).
\end{align}
Upon expanding the plane wave into a series of spherical waves $e^{-i\vec p\cdot\vec r}=4\pi\sum_l\sum_m(-i)^lj_l(pr)Y_{lm}(\theta,\phi)Y_{lm}(\theta\,',\phi\,')$ (primed angels for $\vec p$, and unprimed for $\vec r$) and using the orthogonality relation for the spherical harmonics $\int\int {\rm sin}\theta d\theta d\phi Y_{lm}^*(\theta,\phi)Y_{l\,'m\,'}(\theta,\phi)=\delta_{ll\,'}\delta_{mm\,'}$, plus
a recursion relation for the spherical Bessel functions $\frac{d}{dx}(x^2j_1(x))=x^2j_0(x)$, the matrix element
\begin{align}\label{E1matrixelement}
&<(c\bar{c})_8,\vec p|\vec r|J/\psi> \nonumber\\
=&(-4i)\sqrt{\frac{\pi}{a^3V}}\nabla_{\vec p}[\frac{1}{pa}\int_0^\infty dr r^2j_1(pr)e^{-r/a}] \nonumber \\
=&2^5i\sqrt{\frac{\pi a^3}{V}}\frac{a^2}{(p^2a^2+1)^3}\vec p.
\end{align}
The second line of Eq.~(\ref{E1matrixelement}) indicates that only the $l=1$, i.e., $p$-wave of the final state $(c\bar{c})_8$ ($^3P_J^{(8)}$) survives, implying the $\Delta l=1$ selection rule for the $E_1$ transition (and the change in total spin $\Delta S=0$ since the $H_{E1}$ involves no spin). Substituting Eq.~(\ref{E1matrixelement}) into Eq.~(\ref{E1crosssection}) and working out the integration involving a $\delta$-function, one finally gets the cross section for the gluo-dissociation of $J/\psi$ under Couloumb approximation
\begin{equation}\label{E1JpsiCoulombcrosssection}
\sigma_{E_1, \rm Coulomb}^{g+J/\psi\rightarrow c+\bar{c}}(E_g)=\frac{2^7}{9}g_s^2\frac{\epsilon_B^{5/2}}{m_Q}\frac{(E_g-\epsilon_B)^{3/2}}{E_g^5},
\end{equation}
which is exactly the result of Peskin~\cite{Peskin:1979va}. To arrive at Eq.~(\ref{E1JpsiCoulombcrosssection}), the relation between the Coulombic binding energy of the $1S$ state and the Bohr radius $a^2=1/(m_Q\epsilon_B)$ has been used. In the same way, the cross sections of the gluo-dissociation of the $2S$ state $\Psi\,'$ and the $2P$ state $\chi_c$ under Coulomb approximation can be analytically derived
\begin{equation}\label{E1psi2SCoulombcrosssection}
\sigma_{E_1, \rm Coulomb}^{g+\Psi\,'\rightarrow c+\bar{c}}(E_g)=\frac{2^9}{9}g_s^2\frac{\epsilon_B^{5/2}}{E_g^7m_Q}(E_g-\epsilon_B)^{3/2}(3\epsilon_B-E_g)^2,
\end{equation}
and
\begin{equation}\label{E1kaicCoulombcrosssection}
\sigma_{E_1, \rm Coulomb}^{g+\chi_c\rightarrow c+\bar{c}}(E_g)=\frac{2^7}{9}g_s^2\frac{\epsilon_B^{7/2}}{E_g^7m_Q}(E_g-\epsilon_B)^{1/2}(9E_g^2-20E_g\epsilon_B+12\epsilon_B^2).
\end{equation}

\subsection{Deriving the Gluo-dissociation Cross Sections from $M_1$ transition}
\label{ssec_M1}
Now we turn to deriving the cross sections of gluo-dissociation of heavy quark bound states due to $M_1$ transition. The pertinent interaction Hamiltonian reads
\begin{align}~\label{HM1}
H_{M1}&=-\vec m^a\cdot\vec B^a(t,\vec 0)\nonumber\\
&=-\frac{g_s}{2m_Q}(\frac{\lambda^a}{2}-\frac{\bar{\lambda}^a}{2})(\frac{\vec \sigma}{2}-\frac{\vec \sigma\,'}{2})\cdot \nabla \times \vec A^a(t,\vec 0),
\end{align}
where we have neglected the nonlinear term in the color-magnetic field ($-\frac{1}{2}g_sf^{abc}\vec A^b\times \vec A^c$) that involves two gluon operators and thus does not contribute to the leading order result considered here. Substituting the expansion of the gauge field Eq.~(\ref{Aa}) into Eq.~(\ref{HM1}),
the $M_1$ transition matrix element for the gluo-dissociation process $g(\vec k\lambda)+J/\psi\rightarrow c+\bar{c}$ reads
\begin{equation}
H^{M_1}_{fi}=\frac{ig_s}{2m_Q}\sqrt{\frac{1}{3V\omega_{\vec k}}}<(c\bar{c})_8|(\frac{\vec\sigma}{2}-\frac{\vec\sigma\,'}{2})\cdot(\vec k\times \vec \epsilon_{\vec k\lambda})|J/\psi>,
\end{equation}
where $(\frac{\lambda^a}{2}-\frac{\bar{\lambda}^a}{2})|\rm singlet>=\sqrt{\frac{2}{N_c}}|\rm octet,a>$ has been applied. To make the average over the initial state gluon's polarizations clear, we define a vector $\vec \rho=<(c\bar{c})_8|(\frac{\vec\sigma}{2}-\frac{\vec\sigma\,'}{2})|J/\psi>$ and let it align with the $z$-axis. Then one has $\vec k=(k \rm sin\theta \rm cos\phi,k \rm sin\theta \rm sin \phi, \rm k cos\theta)$ and two physical polarization vectors $\vec \epsilon_1=(\rm cos\theta \rm cos\phi, \rm cos\theta \rm sin\phi, -\rm sin\theta)$ and $\vec \epsilon_2=(-\rm sin\phi,\rm cos\phi,0)$. Furthermore, one notes that for properly defined spin-triplet basis, $(\frac{\sigma_i}{2}-\frac{\sigma\,'_i}{2})|\rm singlet>=|\rm triplet>$ ($i=x,y,z$), which requires that the final state color-octet be a spin-singlet state and the selection rule for the $M_1$ transition be $\Delta S=1$ for the total spin (and $\Delta l=0$ so that the $(c\bar{c})_8$ is a $^1S_0^{(8)}$ state). This also leads to $\rho_i=<(c\bar{c})_8|J/\psi>$ (the inner-product now involves only operations on spatial wave functions). Combining all these observations together, the average over the initial state gluon's polarizations and its incident directions can be worked out to yield the cross section
\begin{eqnarray}\label{M1crosssection}
\sigma_{M_1}^{g+J/\psi\rightarrow c+\bar{c}}(E_g)=\frac{g_s^2\pi}{2\cdot 3}\frac{E_g}{m_Q^2}\frac{V}{(2\pi)^3}\int d^3\vec p \nonumber\\
\times|<(c\bar{c})_8|J/\psi>|^2\delta(E_g-\epsilon_B-\frac{\vec p^2}{m_Q}).
\end{eqnarray}

Using Coulomb wave functions for the $1S$ $J/\psi$ together with the same techniques as in Sec.~\ref{ssec_E1} to handle the octet plane wave function (now only the $l=0$ spherical Bessel function $j_0(pr)$, i.e., the s-wave survives in the spherical wave expansion), one obtains the analytical result for the cross section
\begin{equation}\label{M1JpsiCoulombcrosssection}
\sigma_{M_1, \rm Coulomb}^{g+J/\psi\rightarrow c+\bar{c}}(E_g)=\frac{2^3}{3}g_s^2\frac{\epsilon_B^{5/2}}{m_Q^2}\frac{(E_g-\epsilon_B)^{1/2}}{E_g^3}.
\end{equation}
Similarly, for the $2S$ state $\Psi\,'$ and the $2P$ state $\chi_c$ one has
\begin{equation}\label{M1psi2SCoulombcrosssection}
\sigma_{M_1, \rm Coulomb}^{g+\Psi\,'\rightarrow c+\bar{c}}(E_g)=\frac{2^5}{3}g_s^2\frac{\epsilon_B^{5/2}}{E_g^5m_Q^2}(E_g-\epsilon_B)^{1/2}(E_g-2\epsilon_B)^2,
\end{equation}
and
\begin{equation}\label{M1kaicCoulombcrosssection}
\sigma_{M_1, \rm Coulomb}^{g+\chi_c\rightarrow c+\bar{c}}(E_g)=\frac{2^7}{9}g_s^2\frac{\epsilon_B^{7/2}}{E_g^5m_Q^2}(E_g-\epsilon_B)^{3/2}.
\end{equation}

\section{Gluo-dissociation in an In-medium Potential Model for Various Heavy Quarkonia}
\label{sec_cross-sections}

\subsection{Gluo-dissociation Cross Sections}
\label{ssec_disscrosssections}

Potential models~\cite{Mocsy:2007jz, Riek:2010fk} have been employed to study the fate of heavy quarkonium in the QGP, along with computations of the pertinent spectral functions on the lattice~\cite{Mocsy:2013syh,Burnier:2015tda}. Circumventing the ambiguity in the choice of the appropriate potential (free energy or internal energy of the $Q\bar{Q}$ system), we use, for our purpose of going beyond the Coulomb approximation to calculate the gluo-dissociation of heavy quarkonium, the temperature-dependent potential parameterized in~\cite{Karsch:1987pv} that proved to yield satisfactory description of the vacuum properties (masses, binding energies) as well as reasonable in-medium behavior of various heavy quarkonia below threshold.

The Cornell potential in the vacuum $V(r,0)=-\alpha/r+\sigma r$, with parameters $\alpha=0.471$, $\sigma=0.192\,\rm GeV^2$ and heavy quark masses $m_c=1.320$\,GeV,
$m_b=4.746$\,GeV to reproduce the masses of charmonia and bottomonia below threshold, is modified by color screening in the QGP to become~\cite{Karsch:1987pv}
temperature ($T$) dependent:
\begin{equation}\label{fullpotential}
V(r,T)=-\frac{\alpha}{r}e^{-m_D(T)r}+\frac{\sigma}{m_D(T)}(1-e^{-m_D(T)r}),
\end{equation}
where $m_D(T)$ is the Debye screening mass, for which we take the functional form $m_D/T=-4.058 + 6.32\times(T/T_c-0.885 )^{0.1035}$ fitted to lattice data from~\cite{Burnier:2015tda}, with $T_c=172.5$\,MeV. we are not tempted to include the imaginary part of the in-medium heavy quark potential~\cite{Brambilla:2013dpa,Laine:2006ns,Beraudo:2007ky,Burnier:2014ssa,Liu:2015ypa}, which represents another dissociation mechanism, namely, the Landau damping (similar to the ``quasi-free" first proposed in~\cite{Grandchamp:2001pf}) that arises from the energy transfer from the space-like gluons that mediate interactions between $Q$ and $\bar{Q}$ to the particles in the QGP, and may be closely related to the scattering of the single heavy quark with medium particles~\cite{Laine:2006ns}.

With the heavy quark potential $V(r,T)$ specified above, the Hamiltonian of the non-relativistic $Q\bar{Q}$ system reads $H(r,T)=2m_Q-\nabla^2/m_Q+V(r,T)$, and the eigen-energies and wave functions of various heavy quarkonia are solved from the Schr\"{o}dinger equation~\cite{Karsch:1987pv}
\begin{equation}
[H(r,T)-E_{n,l}(T)]\psi_{n,l}(r,\theta,\phi)=0.
\end{equation}
The temperature-dependent binding energy is then obtained from $\epsilon_B(T)=2m_Q+\sigma/m_D(T)-E_{n,l}(T)$, whose zero point defines the dissociation temperature of the bound state under consideration~\cite{Karsch:1987pv}. The thus solved temperature-dependent wave functions and binding energies are then substituted into Eqs.~(\ref{E1crosssection}) and (\ref{M1crosssection}), to compute the gluo-dissociation cross sections for various heavy quarkonia.

\begin{figure} [!t]
\includegraphics[width=1.05\columnwidth]{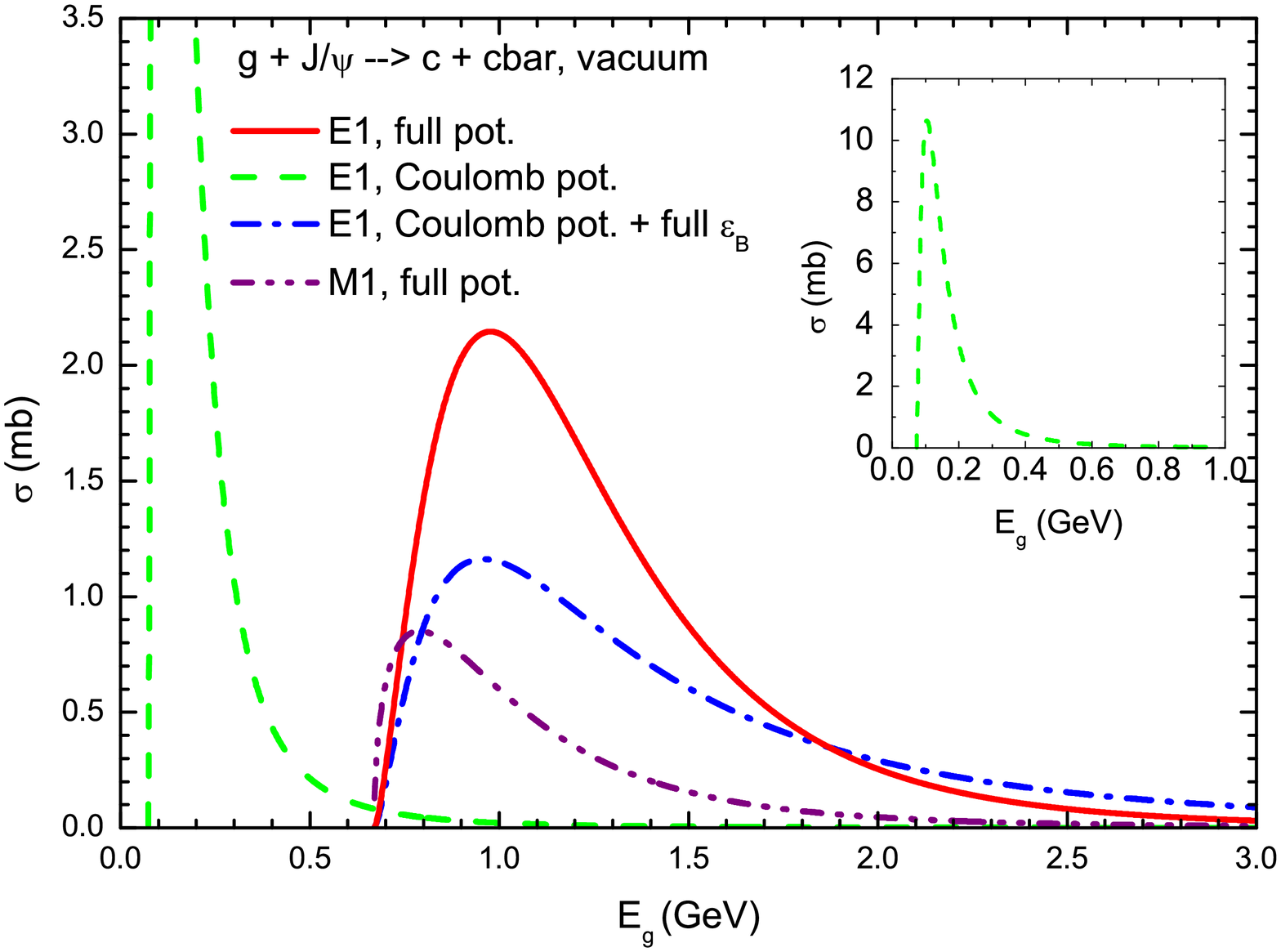}
\vspace{-0.3cm}
\includegraphics[width=1.05\columnwidth]{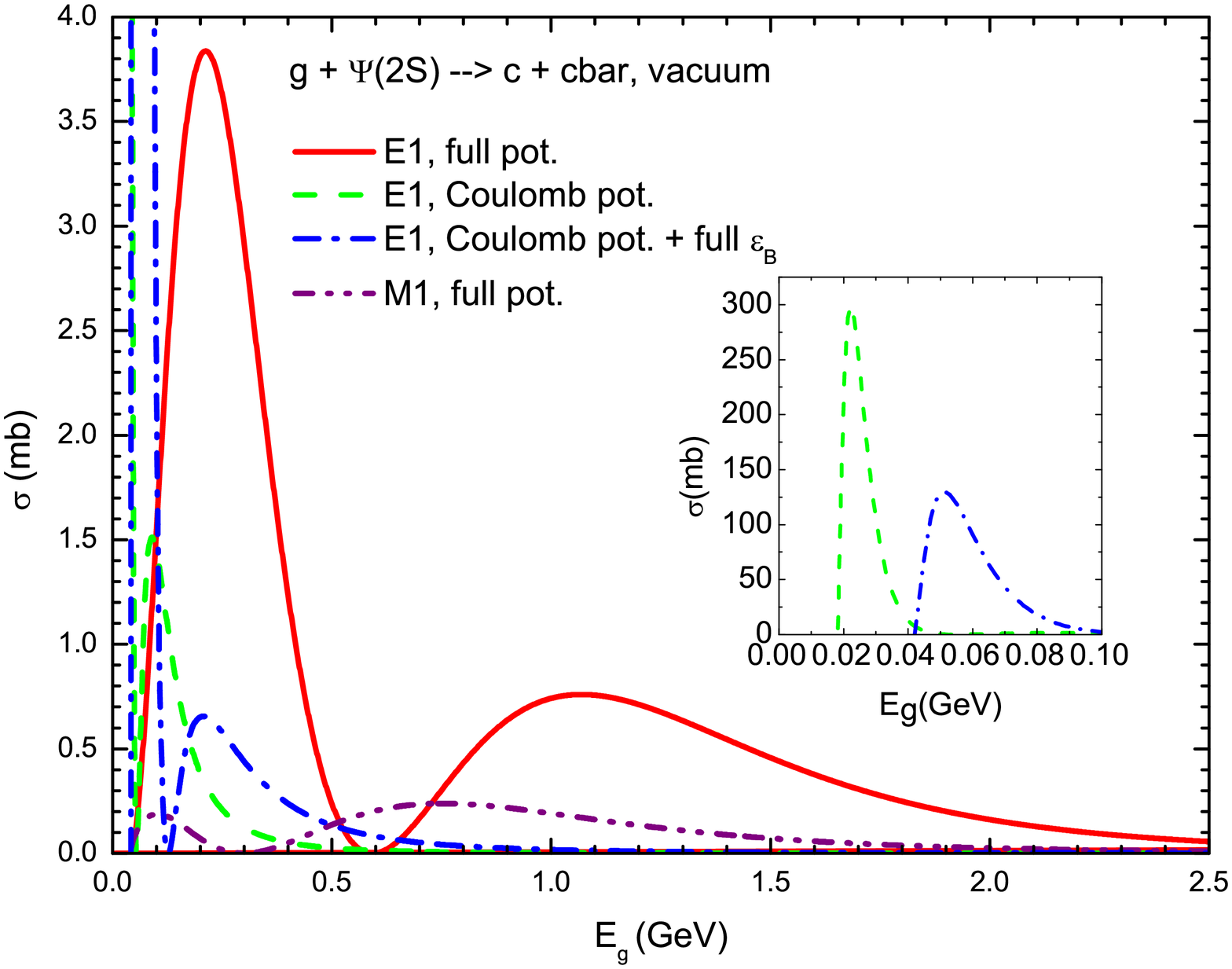}
\vspace{-0.3cm}
\includegraphics[width=1.05\columnwidth]{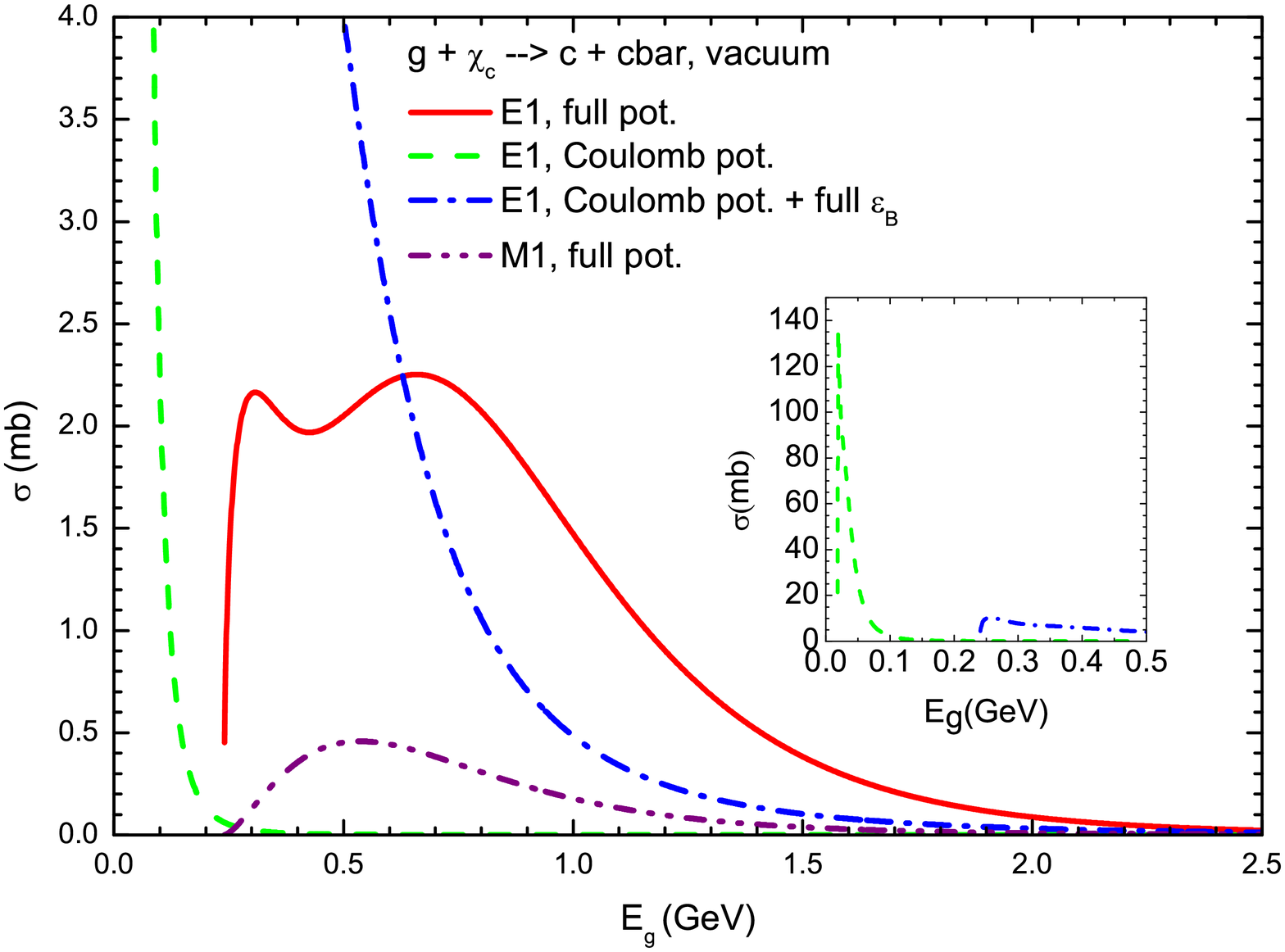}
\vspace{-0.3cm}

\caption{(Color online) Gluo-dissociation cross sections for $J/\psi$ (upper panel), $\Psi(2S)$ (middle panel) and $\chi_c$ (lower panel) in vacuum, respectively. In each case, the $E1$ cross sections from using Coulomb potentials~(green), Coulomb potentials + realistic binding energies~(blue), and full potentials~(red), and the $M1$ cross sections~(purple) are separately displayed. The insertions indicate the pertinent cross sections at full scales.}
\label{fig_vacuumcrosssections}
\end{figure}

Fig.~\ref{fig_vacuumcrosssections} displays the gluo-dissociation cross sections of $J/\psi$, $\Psi(2S)$ and $\chi_c$ in vacuum, respectively, from both $E_1$ and $M_1$ transitions. For each particle, from Coulomb potential result to full potential result, the peak of the $E1$ cross section is much reduced but the strength is distributed over a much broader energy region (the shift of the location to larger gluon energies is simply due to the larger
binding energy that the incident gluon first has to overcome). The broadening of the cross section could be understood from the fact that, the uncertainty in the gluon energy (i.e., the width of the cross section) is roughly of the order of the kinetic energy of the heavy quark in the bound state, which in turn scales inversely with the radius of the bound state; and the bound state is indeed more compact due to further attraction from the confining term in the full potential case. In phenomenological studies, the gluo-dissociation cross sections have been usually taken as the analytical results from using Coulomb potential but with realistic binding energies from full potential calculations~\cite{Rapp:2008tf,Zhou:2014kka}. This ``mixed" result differs substantially ($\sim 50\%$ for $J/\psi$) from the full potential result as indicated in the figure. The $M_1$ gluo-dissociation cross section turns out to be most prominent for $J/\psi$ that has smallest size and overtake the $E1$ cross section at very low energies close to the threshold. The latter is because the $s$-wave scattering for the $J/\psi$ as stipulated by the $\Delta l=0$ selection rule of the $M1$ transition dominates the low energy scattering.

\begin{figure} [!t]
\includegraphics[width=1.05\columnwidth]{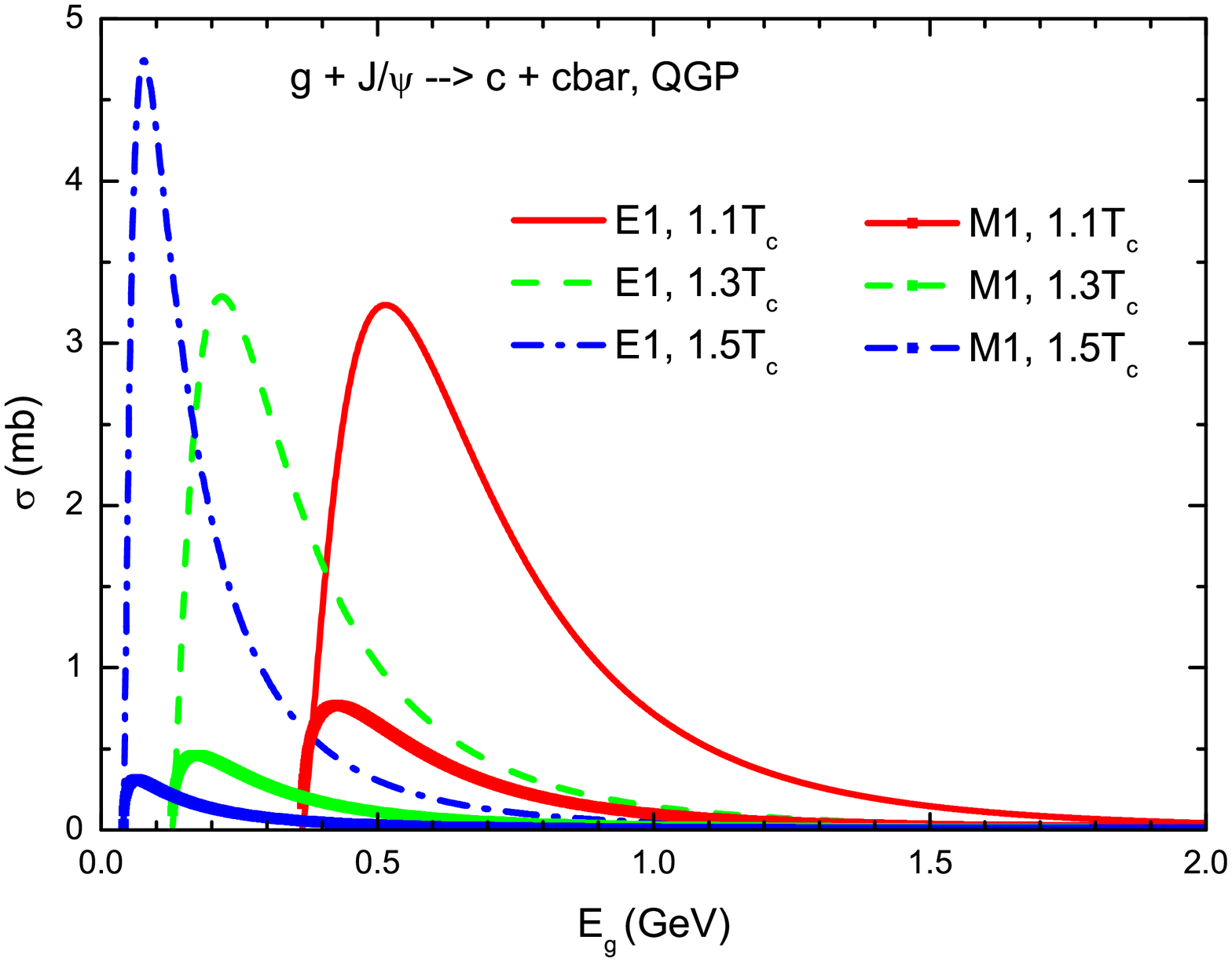}
\vspace{-0.3cm}
\includegraphics[width=1.05\columnwidth]{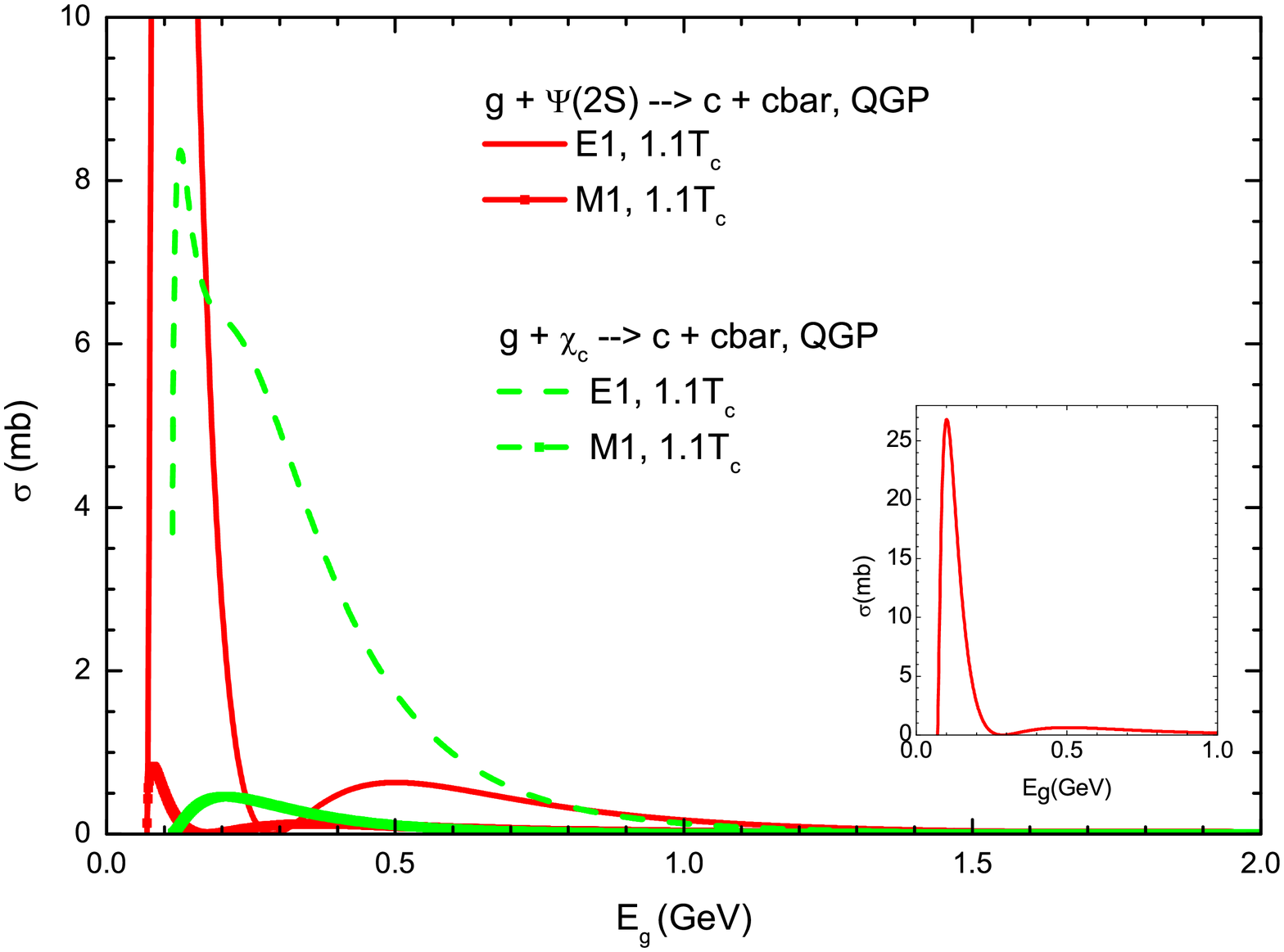}
\vspace{-0.3cm}

\caption{(Color online) Gluo-dissociation cross sections from full potential calculations for $J/\psi$ (upper panel) and $\Psi(2S)$ and $\chi_c$ (lower panel) in QGP at different temperatures up to their respective dissociation temperature. The $E1$ and $M1$ cross sections are displayed separately.
}
\label{fig_QGPcharmoniacrosssections}
\end{figure}

\begin{figure} [!t]
\includegraphics[width=1.05\columnwidth]{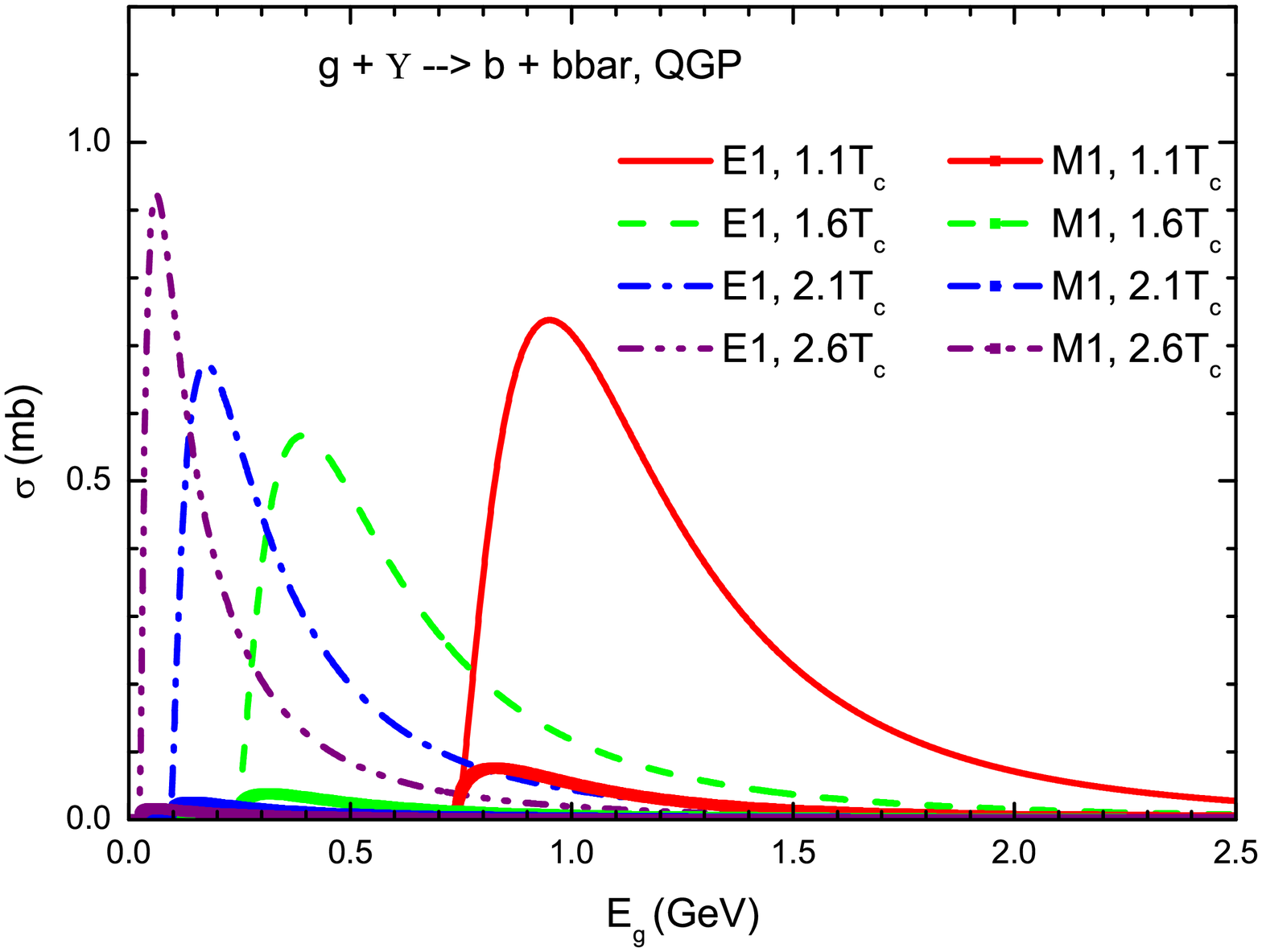}
\vspace{-0.3cm}
\includegraphics[width=1.05\columnwidth]{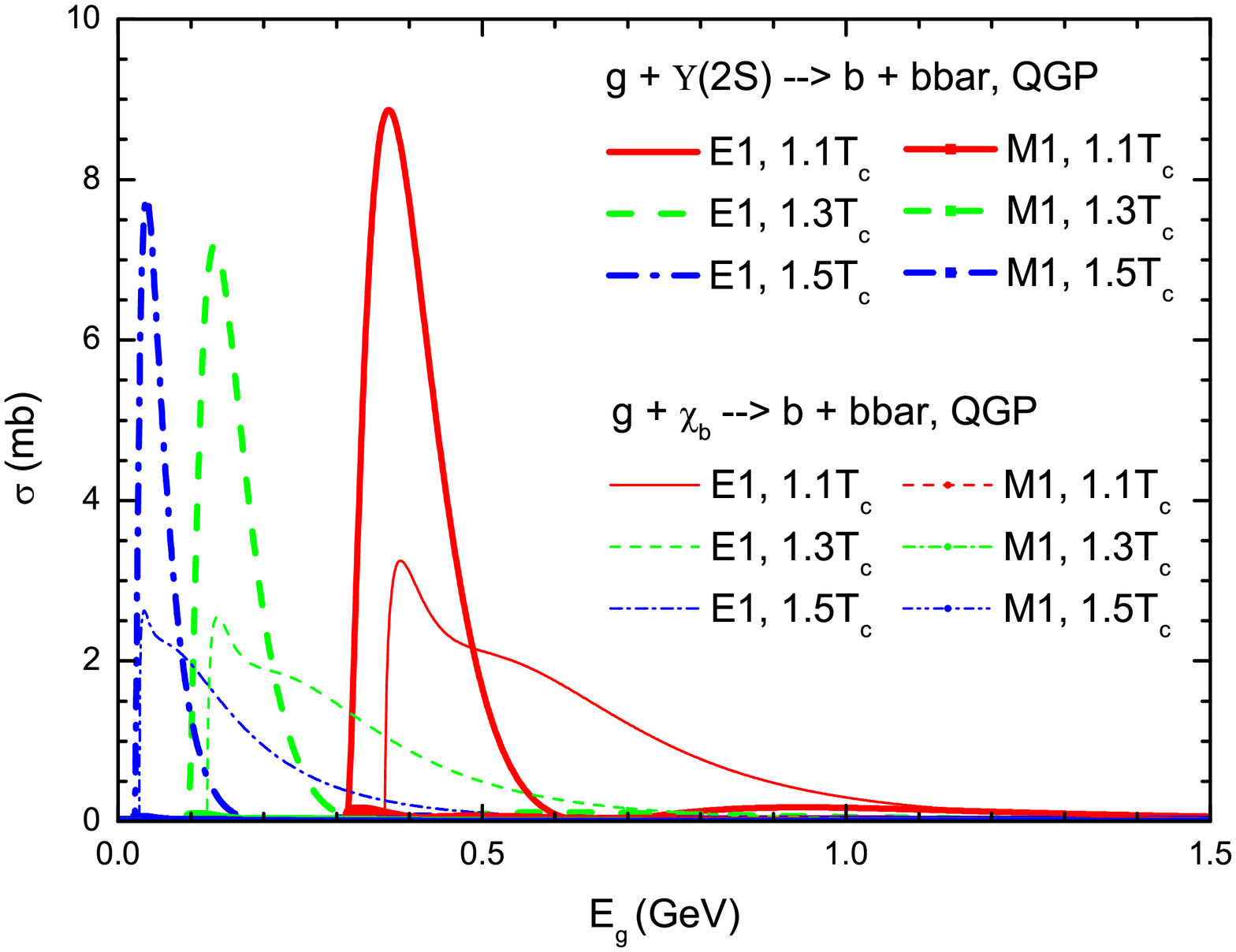}
\vspace{-0.3cm}
\caption{(Color online) The same as Fig.~\ref{fig_QGPcharmoniacrosssections} but for bottomonia.}
\label{fig_QGPbottomoniacrosssections}
\end{figure}

The gluo-dissociation cross sections of various charmonia in the QGP from full potential calculations are summarized in Fig.~\ref{fig_QGPcharmoniacrosssections} up to their respective dissociation temperatures. Again, the $M1$ cross sections are most prominent for $J/\psi$ and decrease with increasing temperature. The latter is due to the fact that as temperature increases, the bound state wave function broadens and extends to the oscillating region of the $s$-wave $j_0(pr)$ of the final color-octet state, resulting in more and more cancellations in the inner product. This is finally over-counteracted, however, by the growing color-electric dipole size in the case of $E1$ matrix element squared; as a result, the temperature dependence of the $E1$ cross section is finally reversed toward higher temperatures. We note that the temperature-dependence of the $E1$ cross section revealed by the full calculations here does not support the simple geometrical scaling (namely obtaining the finite-temperature cross section from the vacuum counterpart by a radius-squared scaling) sometimes adopted in phenomenological studies. The more tightly bound bottomonium states possess significantly higher dissociation temperatures than the corresponding charmonium states and generally much smaller gluo-dissociation cross sections because of smaller size (for $E1$ cross section) and larger quark mass (for $M1$ cross section), c.f., Fig.~\ref{fig_QGPbottomoniacrosssections}. Analogous to the charmonium case, the $M1$ cross section is most prominent for $\Upsilon$; however the ratio between $M1$ and $E1$ cross sections for $\Upsilon$ does not reach that for $J/\psi$, because the decrease of the bound state radius (by a factor of $\sim 2$ ) does not catch up with the increase of the pertinent quark mass (by a factor of more than $3$).

\subsection{Gluo-dissociation Rates}
\label{ssec_dissrates}

\begin{figure} [!t]
\includegraphics[width=1.05\columnwidth]{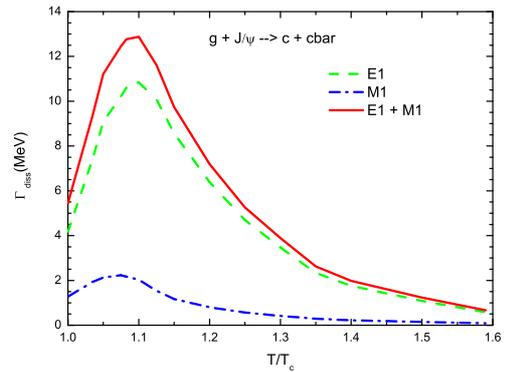}
\vspace{-0.3cm}
\caption{(Color online) Gluo-dissociation rate of $J/\psi$ in the QGP. The $E1$ and $M1$ contributions are displayed separately.}
\label{fig_Jpsi_E1_M1_diss_rate}
\end{figure}
The gluo-dissociation rate which is an input of phenomenological studies~(e.g., \cite{Zhao:2010nk,Zhou:2014kka}) of heavy quarkonia transport in the QGP is obtained by folding the pertinent cross section with the thermal gluon distribution. For a bound state sitting at rest in the QGP, the dissociation rate reads
\begin{equation}
\Gamma_{\rm diss}(T)=d_g\int \frac{d^3k}{(2\pi)^3}f_g(E(\vec k))v_{\rm rel}\sigma(|\vec k|,T),
\end{equation}
where $d_g=2\cdot8=16$ is the gluon degeneracy and $f_g=1/(e^{E(\vec k)/T}-1)$ the Bose distribution with gluon energy $E(\vec k)=\sqrt{\vec k^2 + m_g^2(T)}$. The thermal gluon mass is taken to be
$m_g(T)=\sqrt{3/4}gT$ with fixed $g=2.3$ for $N_f=3$ active light flavors and $N_c=3$ colors~\cite{Riek:2010fk}.

\begin{figure} [!t]
\includegraphics[width=1.05\columnwidth]{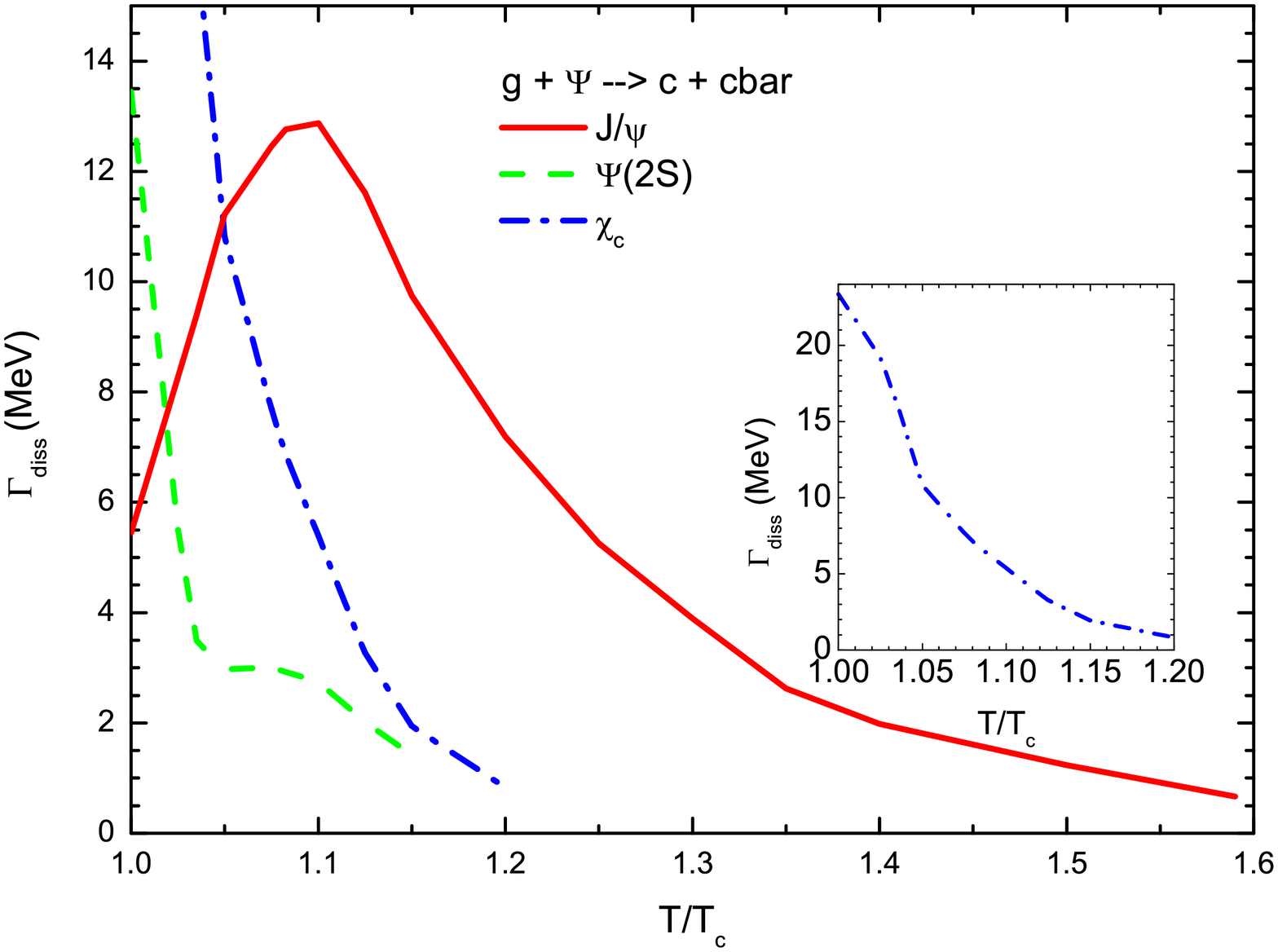}
\vspace{-0.3cm}
\includegraphics[width=1.05\columnwidth]{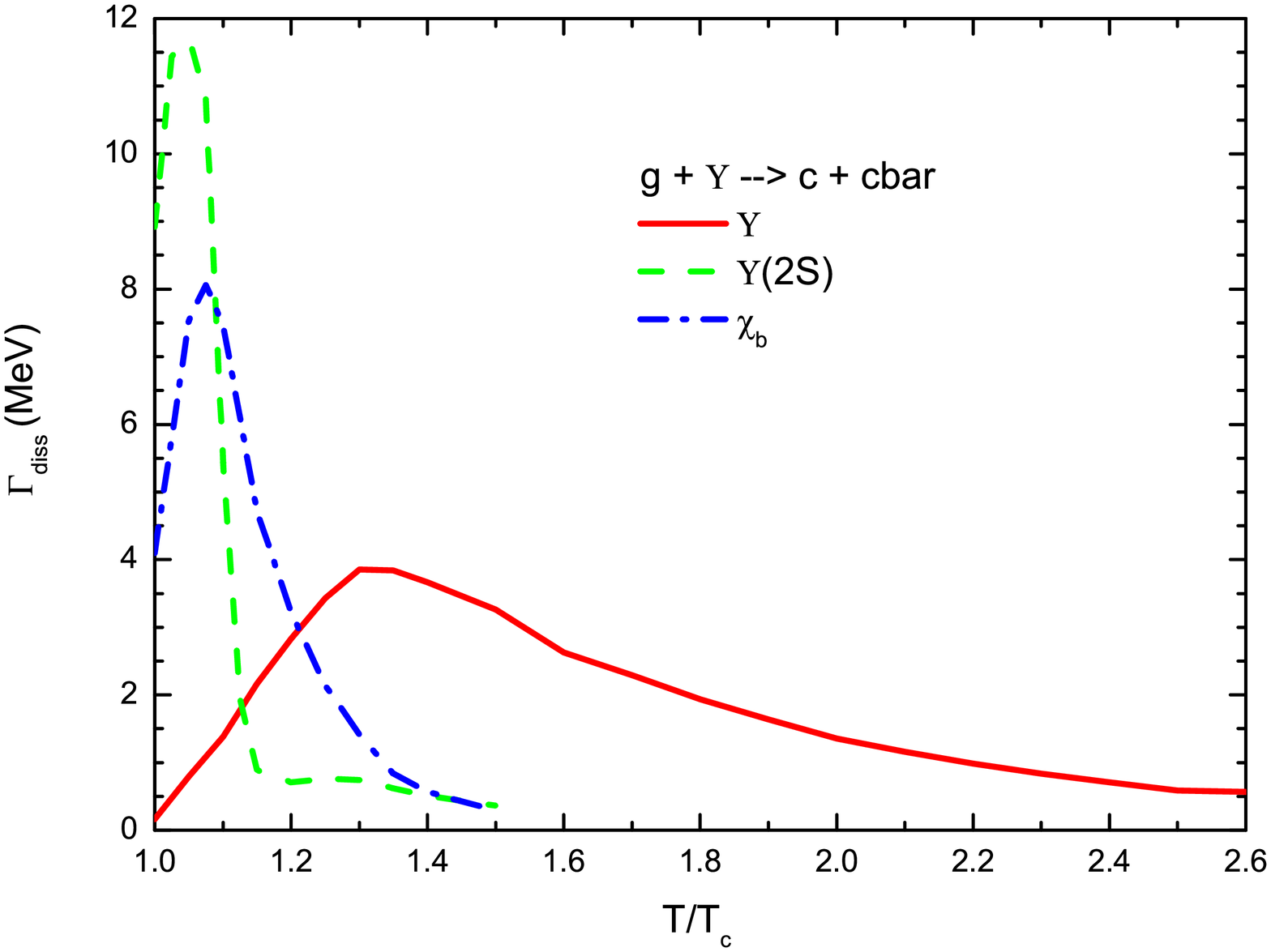}
\vspace{-0.3cm}

\caption{(Color online) Gluo-dissociation rate ($E1+M1$) of charmonia (upper panel) and bottomonia (lower panel) in the QGP up to their respective dissociation temperatures.}
\label{fig_Psi_Upsilon_total_rates}
\end{figure}

The gluo-dissociation rate of $J/\psi$ up to its dissociation temperature is displayed in Fig.~\ref{fig_Jpsi_E1_M1_diss_rate}. In the temperature range from $T_c$ to $1.2T_c$, the $M1$ contribution accounts for $\sim 10\%-25\%$ of the total dissociation rate, in accord with the pronounced $M1$ cross section~(c.f. Fig.~\ref{fig_QGPcharmoniacrosssections}) in this temperature region. Although this is not comparable to the $E1$ contribution, it may still of phenomenological
significance, given that the fireball would spend a significant duration near $T_c$ because of the softening of the equation of state (EoS). For other particles, the $M1$ contribution generally accounts for less than $10\%$ of the total dissociation rate and thus is not seprately shown; see Fig.~\ref{fig_Psi_Upsilon_total_rates}, where the total ($E1+M1$) dissociation rates for various charmonia and bottomonia up to their respective dissociation temperatures are compiled. In general, the dissociation rates of the more tightly bound bottomonia are smaller than those of the charmonia; but that is not the case for $\Upsilon(2S)$ versus $\Psi(2S)$ in the temperature region $T_c-1.1T_c$, as a result of larger phase space overlap of the $\Upsilon(2S)$ cross sections
with thermal gluons.

In previous calculations with Coulomb potential approximation, the gluo-dissociation rate of the $J/\psi$ was found to decrease monotonously with increasing temperature~\cite{Park:2007zza}, due to the fact that as the binding energy gets lower, the gluo-dissociation cross section (shifting toward lower energy) has less and less overlap with the phase-space weighted gluon distribution function $k^2f_g$. However, we find, in the present full potential calculation~(with larger binding energies and broader cross sections than the calculation with Coulomb approximation in~\cite{Park:2007zza}), the dissociation rate of the $J/\psi$ exhibits an increase at low temperatures, peaking at $\sim 1.1T_c$ and then followed by the usual monotonous decrease toward higher temperatures. This increase at low temperatures, also seeable for $\Upsilon(2S)$ and $\chi_b$ but disappearing for the loosely bound $\Psi(2S)$ and $\chi_c$, becomes more pronounced and persists to larger temperature ($\sim 1.3T_c$) for the most tightly bound $\Upsilon$ (the dissociation rate of the $\Upsilon$ nearly vanishes at temperatures close to $T_c$ because the thermal gluon energy is too low compared to its binding energy). Recall that, while yielding to the next-to-leading order counterpart at high temperatures (i.e., small binding energies)~\cite{Grandchamp:2001pf,Park:2007zza}, the gluo-dissociation as the leading order inelastic break-up mechanism of the heavy quarkonia should dominate over the former, as soon as the bound state is still sufficiently tightly bound (i.e., at low temperatures) such that the incident gluon of long wavelength does not resolve the substructure of the bound state. Therefore, it is not surprising that the calculated dissociation rates of $J/\psi$ and $\Upsilon$ demonstrate an increase with increasing thermal gluon energy in the temperature region where they are still sufficiently tightly bound and thus the gluo-dissociation dominates the break-up of the bound states.

\section{Summary}
\label{sec_sum}
In this work, using an effective Hamiltonian derived from the QCD multipole expansion, we have calculated the gluo-dissociation cross sections of various heavy quarkonia in the QGP. While the color-electric dipole ($E1$) transition allows us to reproduce Peskin's result from OPE analysis in the Coulomb potential approximation, the color-magnetic dipole ($M1$) transition considered here for the first time has shown to be significant at low energies close to the scattering threshold in accord with the pertinent selection rules. We have then carried out a full calculation of the gluo-dissociation rates for various charmonia and bottomnia within a non-relativitic in-medium potential model. The $M_1$ contribution has been shown to be most prominent for the $J/\psi$ and account for $\sim 10\%-25\%$ of the total ($E1+M1$) dissociation rate in the temperature range $T_c-1.2T_c$, which may be of phenomenological significance. Furthermore, taking into account of the (screened) confining potential results in an increase of the dissociation rate for tightly bound $J/\psi$ and $\Upsilon$ at low temperatures where the present gluo-dissociation as the leading order mechanism for the break-up of heavy quarkonia in the QGP is mostly applicable.

The dissociation rates calculated here unphysically decrease toward higher temperatures, which is known to be an artifact of the leading order approximation~\cite{Beraudo:2007ky} and should be cured by the next-to-leading order (e.g., $g+J/\psi\rightarrow g + c + \bar{c}$) calculations~\cite{Zhao:2010nk,Brambilla:2013dpa,Grandchamp:2001pf,Park:2007zza}. The latter is expected to take over at high temperatures when the binding energies of bound states become small relative to the Debye screening mass. The present quantum-mechanical perturbation method should be relatively readily extended to the next-to-leading order, with the prospect of making the pertinent calculations more accessible than those in literature~\cite{Brambilla:2013dpa,Park:2007zza}. Such calculations are on-going and the results will be presented in a future work.\\

{\bf Acknowledgments:}
We are indebted to Ralf Rapp for careful reading of the manuscript and useful remarks. This work was supported
by NSFC grant 11675079.

\end{document}